\documentclass{article}

\usepackage{arxiv}

\usepackage[utf8]{inputenc} 
\usepackage[T1]{fontenc}    
\usepackage{hyperref}       
\usepackage{url}            
\usepackage{booktabs}       
\usepackage{amsfonts}       
\usepackage{nicefrac}       
\usepackage{microtype}      
\usepackage{lipsum}		
\usepackage{graphicx}
\usepackage[style=numeric,minnames=6,maxnames=6,backend=biber]{biblatex}
\usepackage{doi}
\usepackage{amsmath}
\usepackage{setspace}

\def \paperTitle {Ensuring accurate stain reproduction in deep generative networks for virtual immunohistochemistry.}

\title{\paperTitle}

\author{ \href{https://orcid.org/0000-0002-6890-2401}{\includegraphics[scale=0.06]{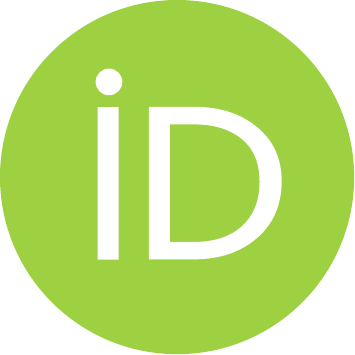}\hspace{1mm}Christopher~D.~Walsh}\thanks{Corresponding author.} \\
	CRUK Beatson Institute,\\
	Glasgow, UK\\
	\texttt{c.walsh.1@research.gla.ac.uk} \\
	\And
	\href{https://orcid.org/0000-0002-7192-6906}{\includegraphics[scale=0.06]{images/orcid.pdf}\hspace{1mm}Joanne~Edwards} \\
	Institute of Cancer Sciences,\\
	University of Glasgow,\\
	Glasgow, UK\\
	\texttt{Joanne.Edwards@glasgow.ac.uk} \\
	\And
	\href{https://orcid.org/0000-0003-4898-040X}{\includegraphics[scale=0.06]{images/orcid.pdf}\hspace{1mm}Robert~H.~Insall} \\
	Institute of Cancer Sciences,\\
	University of Glasgow,\\
	Glasgow, UK\\
	\texttt{Robert.Insall@glasgow.ac.uk} \\
}

\renewcommand{\shorttitle}{A Loss Function for Virtual Immunohistochemistry}

\hypersetup{
pdftitle={\paperTitle},
pdfsubject={Deep Learning, Pathology},
pdfauthor={Christopher D.~Walsh, Robert H.~Insall},
pdfkeywords={Deep Learning, Pathology},
colorlinks=true,       
linkcolor=black,          
citecolor=black,        
filecolor=black,      
urlcolor=black           
}

\usepackage{todonotes}

\begin{document}
\maketitle

\begin{abstract}
    Immunohistochemistry is a valuable diagnostic tool for cancer pathology. However, it requires specialist labs and equipment, is time-intensive, and is difficult to reproduce. Consequently, a long term aim is to provide a digital method of recreating physical immunohistochemical stains. Generative Adversarial Networks have become exceedingly advanced at mapping one image type to another and have shown promise at inferring immunostains from haematoxylin and eosin. However, they have a substantial weakness when used with pathology images as they can fabricate structures that are not present in the original data. CycleGANs can mitigate invented tissue structures in pathology image mapping but have a related disposition to generate areas of inaccurate staining. In this paper, we describe a modification to the loss function of a CycleGAN to improve its mapping ability for pathology images by enforcing realistic stain replication while retaining tissue structure. Our approach improves upon others by considering structure and staining during model training. We evaluated our network using the Fr\'echet Inception distance, coupled with a new technique that we propose to appraise the accuracy of virtual immunohistochemistry. This assesses the overlap between each stain component in the inferred and ground truth images through colour deconvolution, thresholding and the Sorensen-Dice coefficient. Our modified loss function resulted in a Dice coefficient for the virtual stain of 0.78 compared with the real AE1/AE3 slide. This was superior to the unaltered CycleGAN's score of 0.74. Additionally, our loss function improved the Fr\'echet Inception distance for the reconstruction to 74.54 from 76.47. We, therefore, describe an advance in virtual restaining that can extend to other immunostains and tumour types and deliver reproducible, fast and readily accessible immunohistochemistry worldwide.
\end{abstract}


\onehalfspacing

\section{Introduction}
\label{sec:introduction}

Haematoxylin and Eosin (H\&E) is the default stain in pathology and is frequently used for a wide variety of tasks and endures in approaches for their automation, like segmentation of glands in the prostate \cite{li_automated_2020}, classification of early pancreatic cancer \cite{langer_computer-aided_2015} and staging of colorectal cancer \cite{fleming_colorectal_2012}. Deep networks can perform these tasks on H\&E alone, suggesting that a broad scope of information is available from H\&E; however, it indiscriminately stains both tumour and non-tumour cells. Therefore, it can be challenging to diagnose cases with ambiguous histology or poor tumour differentiation with the unaided eye, as humans have difficulty distinguishing the subtle changes in colouring. In these cases, immunohistochemistry (IHC) is employed. It visualises the location and quantities of specific molecules that are present in tissue using an antigen-antibody reaction \cite{kim_immunohistochemistry_2016}. It is widely used in diagnosis, especially in challenging cases, because tumours can express specific antigens that can be highlighted by IHC \cite{duraiyan_applications_2012}. This allows the tumour cells to be differentially distinguished from non-tumour cells.

However, IHC is complex, consumes more tissue, is more expensive than H\&E and is hard to reproduce. This suggests a need for a more accessible, economical and reproducible method to attain an improved level of contrast. Through virtual stain translation, deep learning can accomplish this by converting between H\&E and the desired IHC stain. Progress has already been made in this space, such as a virtual SOX10 stain for melanoma \cite{jackson_machine_2020}, a virtual PAS stain for renal slides \cite{lo_cycle-consistent_2021} and virtual immunofluorescent images \cite{burlingame_shift_2020}.

Deep learning models typically belong to one of two groups. They are either generative or discriminative \cite{ronneberger_u-net_2015}. Discriminative models typically perform tasks like classification or image segmentation. They usually predict the probability that a given input belongs to one class or another \cite{fragemann_review_2022}. The role of generative models is not to predict a class or label but to learn the distribution of the input data and then recreate new examples. Nearly all state-of-the-art approaches to virtual stain translation use some form of deep generative network.

\begin{figure}[h]
    \centering
    \includegraphics[width=0.9\textwidth]{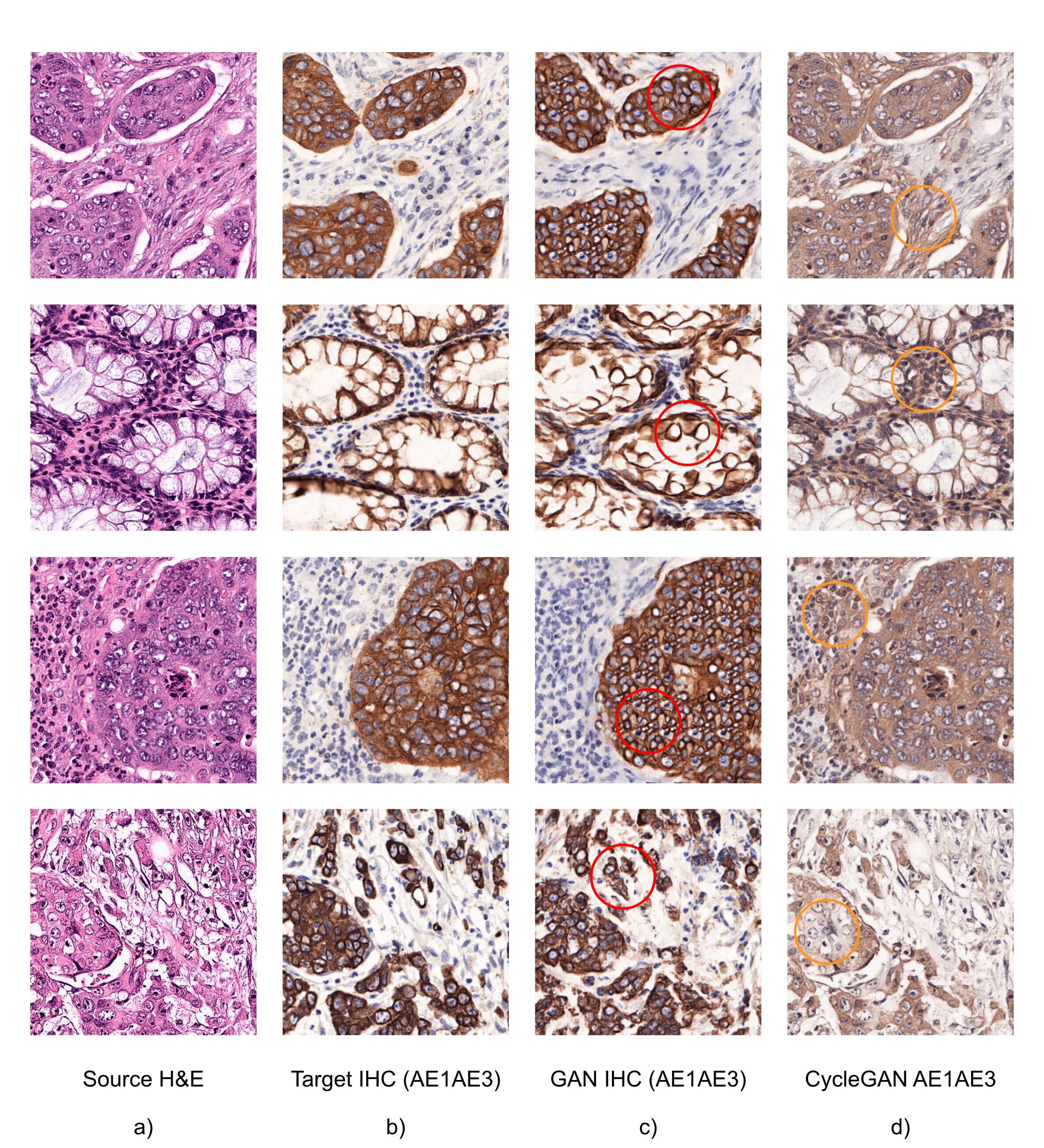}
    \caption{Issues with generative deep networks for virtual immunohistochemistry: a) source tissue, b) target tissue, c) virtual IHC with a GAN. Red circles show invented structural details. d) virtual IHC with a CycleGAN. Orange circles show areas of inaccurate staining.}
    \label{fig:gan_example}
\end{figure}

In recent years Generative Adversarial Networks (GANs) have become particularly popular for image translation\cite{gui_review_2020}. GANs were introduced by Goodfellow et al. in 2014 \cite{goodfellow_generative_2014}. They are network architectures that apply an adversarial loss function during training. They consist of two subnetworks. The first is termed the Generator (G); Its goal is to generate data in the target domain Y, either using input data from a domain X or random noise. The next is the discriminator (D); Its goal is to differentiate between domain Y and generated examples. The generator and discriminator are typically deep networks, but, in theory, they can be any function approximator that is differentiable \cite{gui_review_2020}. The generator is trained to reproduce the input dataset. The discriminator is usually a binary classifier that attempts to distinguish generated examples from the actual ground truth. The two models are trained in competition, aiming for the generator to make synthetic images that the discriminator cannot effectively distinguish from real ones. The training process is essentially a min-max optimisation problem \cite{fragemann_review_2022} and should terminate when the generator's loss has reached a minimum and the discriminator has reached a maximum. By enforcing the comparison between ground truth data and the generated samples, GANs can learn to create new examples of many data types, including images. This capacity for realistic approximation of the input data is powerful and engaging and explains their use in prior virtual staining attempts. However, in pathology, particularly when used for virtual stain translation, this ability to produce a highly realistic image is also perilous. The weakness exists because GANs can potentially fabricate tissue and cell structures that may not exist in the input data but look so natural that the discriminator does not detect them. This is a significant liability for medical use when a diagnosis critically depends on precise and faithful reporting of structural details in the tissue. Figure \ref{fig:gan_example} shows several source H\&E images and the corresponding IHC image from a serial section, along with virtual stains generated by two deep networks. Column c) shows examples where a GAN has recreated the IHC version of the source tissue. Comparing the tissue shown in the red circles with the source tissue in column a) demonstrates that the GAN has wholly changed the tissue structure.

One type of Generative Adversarial Network, a CycleGAN, is designed for image translation \cite{saxena_comparison_2021} and can alleviate this issue. It is trained using an adversarial process involving generators and discriminators, but a CycleGAN has two pairs. The first generator translates an image from the source domain to the target. Its discriminator is guided to become increasingly adept at recognising fakes in the target domain. This forces the generator to produce realistic images as it competes with the discriminator. The next generator in the cycle translates the inferred target image from the first to an image in the source domain. Its discriminator attempts to distinguish this image from the ground truth, again constraining the generator to produce realistic images from the source domain. The final step is that a loss term is added to the network to apply a mean absolute error calculation between the output of the second generator and the source image, thus forcing the network to make them alike. This is known as cycle consistency loss \cite{zhu_unpaired_2020}. For the source and output image to be identical, both generators must maintain the structural detail throughout the process. This ensures that the first generator maintains the structure when translating from the source to the target domain and mitigates the drawback of invented structural details. Column (d) of figure \ref{fig:gan_example} shows examples of a CycleGAN recreating the IHC version of the source tissue. Comparing the tissue structure of the virtual patches with the source tissue in column (a) reveals that it has exactly reproduced the structural detail. However, an unaltered CycleGAN still has no reason to enforce detailed reproduction of the correct staining as long as the image behaves realistically. Comparing the tissue highlighted by the orange circles with the source H\&E in column (a), the structure is identical. However, the stained areas are considerably different from the ground truth IHC image in column (b).

One characteristic of real H\&E and IHC stains is that they can vary significantly in intensity and hue depending on the preparation and age of the stain and the type of light source used when imaging, among many other factors. This variation in stain colour means that colour normalisation is required to ensure the input data matches the training data when deep networks are used to infer data gathered on a different day or from a different lab. Several such methods are used, such as the Macenko method \cite{macenko_method_2009} which uses singular value decomposition in optical density space to estimate vectors that express the stain ratios of each pixel. This vector can then be adjusted to match the desired target. Or Reinhard's method \cite{reinhard_color_2001}, which converts the slide to LAB colour space and then shifts the mean and standard deviation of each channel of the source image to match the target. Another approach is the Vahadane structure-preserving colour normalisation method \cite{vahadane_structure-preserving_2016}, which uses sparse non-negative matrix factorisation to determine a matrix of stain ratios in optical density space. The inverse of this can be computed and used to calculate staining intensity. Converting the RGB image to a stain colour space allows all pixel values to be altered equally in each channel and leaves the relative values unaltered, preserving the structure. Each has its advantages and drawbacks, but whichever is chosen, some process must be used to normalise the stain colour to ensure the slides belong to the same colour domain before use in training or inference. Additionally, implementing a technique to convert from RGB whole slide images to a stain colour space allows for an analysis of the intensity of each virtual stain and the determination of its similarity to the ground truth, which is particularly useful when assessing virtual IHC. This paper introduces a new technique to produce a quantitative value representing the accuracy of a virtual IHC image based on colour deconvolution, intensity thresholding and the Sorensen-Dice coefficient.

We have observed that GANs can produce realistic pathology images with an adequate quantity of correctly normalised input data. However, they have a significant weakness when used for pathology data: the potential to invent tissue. A CycleGAN solves the problem of invented structural details. But fails to ensure accurate reproduction of the tissue colouring. This is the problem for which this paper suggests a solution. We propose an amendment to the CycleGAN loss function where the goal is to balance the network's focus between stain and structural reproduction and allow for the creation of reliable virtual IHC images. This can deliver fast, affordable and more readily accessible immunohistochemistry, using a process that could be extended to many types of cancer and stains.

\newpage
\section{Materials and Methods}

\subsection{Dataset}

To train a neural network to translate from H\&E to our chosen stain, we required a ground truth dataset of structurally paired H\&E and AE1/AE3 slides. There was no available dataset with these attributes; therefore, we generated one. We did this by cutting serial sections at a close separation to ensure the tissue was congruent between slices with minimal deformation. We then stained the first slice with H\&E and the second with our desired IHC stain pan-cytokeratin AE1/AE3.

The Glasgow Tissue Research Facility supplied the source tissue along with the NHSGGC Biorepository (ethics no 22/ws/0207). They fully anonymised the slides and metadata before use in this study. The Edwards group in the Institute of Cancer Sciences at the University of Glasgow cut pairs of serial sections at 2.5-micron intervals from eight CRC resections, resulting in sixteen total sides. The first section of each pair was stained with H\&E and the second with pan-cytokeratin AE1/AE3. The Edwards group and histology services at the Beatson Institute for Cancer Research carried out the AE1/AE3 IHC staining. Prepared slides were scanned with a Hamamatsu S60 slide scanner, with an LED light source, at 20x magnification.

A test dataset was also required to validate the virtual IHC slides and compare model performance. Therefore, we had two additional slides cut and stained using the same process as above but performed on a different day to ensure the stain preparation was distinct; this effectively simulated real variations in staining. The Glasgow Tissue Research Facility cut the sections and carried out H\&E staining, and members of the Edwards group performed the AE1/AE3 IHC staining. We then carried out post-processing steps in the same manner as the primary dataset, and we reserved these slides as unseen test data for network evaluation and comparison.

Our raw dataset had not yet satisfied two quality and accuracy requirements for training a deep network for virtual IHC. The first was that the H\&E and AE1/AE3 tissue structure had to be identical between sections, matched right down to the cellular level; The second was that the training, validation and test input data all had to belong to the same colour domain as the training data. This meant that we had to address the variation in colour profile across slides.

We settled on using the Vahadane stain normalisation method to standardise the colour profile across slides. We amended this method in several ways; these were mainly to improve reliability and avoid slide artefacts affecting the resulting stain matrix. Our first modification to the technique was to sample pixels across the entire slide to compute our stain matrix rather than just a few points. We created a tissue mask by thresholding a low-resolution image of the slide in LAB colourspace. We could then subsample points from all of the tissue, and it became more reliable to compute the stain matrix on slides with unusual artefacts. The next and most significant difference in our application of this method is that we used non-negative matrix factorisation rather than sparse non-negative factorisation. The original Vahadane method employed sparse factorisation as they speculated that a pixel is composed of exclusively one stain or the other. However, we propose that it more accurately reflects the true biology if a ratio of the stains expresses the light absorbed by the tissue at any one pixel.

We applied this modified Vahadane stain normalisation method to our raw slide dataset and finally carried out image registration and alignment using a mutual information-based image similarity metric. We shall describe this process in detail in another paper. This pipeline provided a normalised and aligned set of paired source and target patches to train our deep networks. This dataset can be made available upon request.

\subsection{Methods}

\subsubsection{GAN based Virtual Immunohistochemistry}

For use as a reference in comparison with our model, we implemented a virtual IHC stain translation network based on a Generative Adversarial Network (GAN) as described in section \ref{sec:introduction}. The details of the GAN loss function are given in equation \ref{eqn:gan-adversarial-loss}. It demonstrates that we trained the discriminator D to maximise the probability of assigning the correct label to both real examples and samples from G. We simultaneously train the generator G to minimise $log(1-D(G(z)))$ to fool the discriminator into assigning the real label to generated images. The discriminator must learn to make its output as realistic as possible.

\begin{equation}
    \mathcal{L}_{GAN}(G,D_Y,X,Y) = \mathbb{E}_{y~p_{data}(y)}[log D_Y(y)] + \mathbb{E}_{x~p_{data}(x)}[log(1 - D_Y(G(x))]
    \label{eqn:gan-adversarial-loss}
\end{equation}

\subsubsection{CycleGAN for Virtual Immunohistochemistry}

For use as a control, we implemented an unaltered CycleGAN as is described in the original paper from 2017 by Zhu et al. \cite{zhu_unpaired_2020}. The architecture of the network and training method is as described in this paper and section \ref{sec:introduction}. The details of the Cycle Consistency Loss proposed by Zhu et al. are given in equation \ref{eqn:gan-cycle-loss}, where G is the generator from the first pair and F is the generator from the second pair.

\begin{equation}
    \mathcal{L}_{cyc}(G,F) = \mathbb{E}_{x~p_{data}(x)}[\left\Vert F(G(x))-x \right\Vert_1] + \mathbb{E}_{y~p_{data}(y)}[\left\Vert G(F(y))-y \right\Vert_1]
    \label{eqn:gan-cycle-loss}
\end{equation}

\subsubsection{Mid-cycle loss CycleGAN for Virtual Immunohistochemistry}

Our modification to the unaltered CycleGANs loss function was that we added a term to minimise the differences between the generated image and the real image at the midpoint of the cycle, as is shown in equation \ref{eqn:gan-mid-loss}. We refer to this term as the ``mid-cycle loss''. In the unaltered CycleGAN, the cycle loss is multiplied by a scaling factor to weight it with regard to the adversarial loss, usually called lambda 1\cite{zhu_unpaired_2020}. We also added a weighting coefficient to the mid-cycle loss, lambda 2.

\begin{equation}
    \mathcal{L}_{midcyc}(G) = \mathbb{E}_{x,y}[\left\Vert G(x) - y \right\Vert_1]
    \label{eqn:gan-mid-loss}
\end{equation}

We propose that mid-cycle loss will balance the network and maintain the desired stain. Simultaneously, the cycle consistency loss should enforce the structure, and the adversarial loss ensures the images be as realistic as possible.

\begin{figure}[h]
    \centering
    \includegraphics[width=\textwidth]{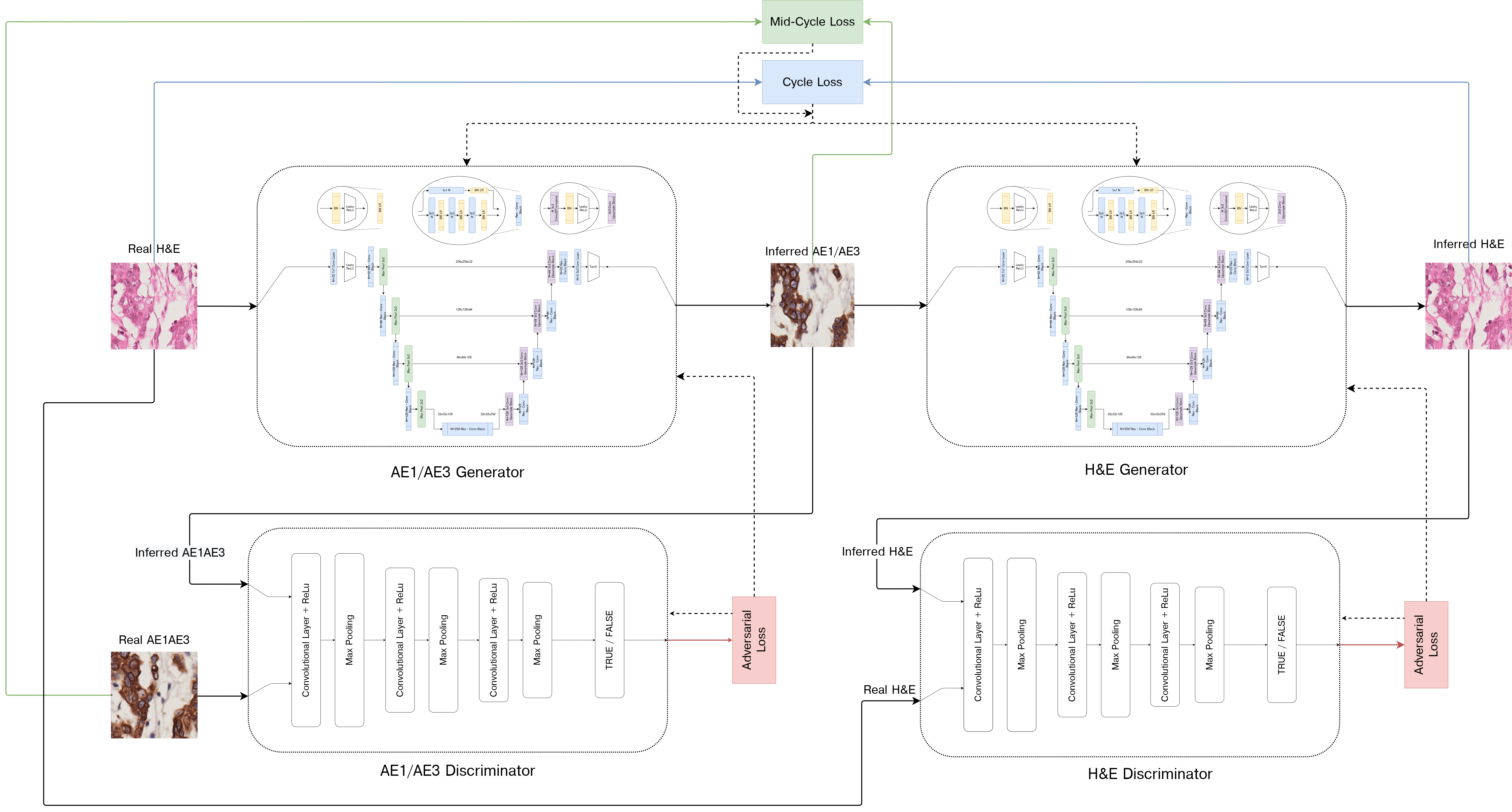}
    \caption{CycleGAN Architecture for Virtual Immunohistochemistry}
    \label{fig:cycleganArch}
\end{figure}

Our network architecture for the virtual IHC CycleGAN with mid-cycle loss is given in figure \ref{fig:cycleganArch}. Displayed within are the AE1/AE3 Generator (G$_{AE}$) and AE1/AE3 Discriminator (D$_{AE}$), and the H\&E Generator and Discriminator (G$_{HE}$) and (D$_{HE}$). The generators and discriminators are configured to train using adversarial loss, and the two generators are linked for Cycle Consistency Loss. We used an encoder network with nine 2D convolutional layers followed by batch normalisation and leaky rectified linear units for the generators. The output feature maps halved in dimensionality after each layer going from 256x256x64 to 1x1x512 by the time the latent space was reached, starting with 64 filters per layer increasing by doubling until a limit of 512 was reached. The decoder reversed this pattern restoring the original dimensions using nine 2D Convolutional Transpose Layers, this time halving the number of filters at each layer, ranging from 512 to a lower limit of 64 by the second last layer. The last layer of the decoder was a Convolutional2D transpose layer with a tanh activation function and three output filters to reproduce the desired RGB image.

The discriminator networks consisted of three sets of convolutional, batch normalisation and rectified linear unit layers followed by a dense layer for probability output. A modified loss function was used in the final network with our mid-cycle loss term. The function is a combination of the adversarial loss (equation \ref{eqn:gan-adversarial-loss}), cycle consistency loss (equation \ref{eqn:gan-cycle-loss}) and our mid-cycle loss (equation \ref{eqn:gan-mid-loss}). It is given in equation \ref{eqn:cyclepath-loss}:

\begin{equation}
    \begin{split}
        \mathcal{L}(G_{AE},G_{HE},D_{AE},D_{HE}) & = \mathcal{L}_{GAN}(G_{AE},D_{AE},X,Y) \\ & + \mathcal{L}_{GAN}(G_{HE},D_{HE},X,Y) \\ &+ \lambda_1 \mathcal{L}_{cyc}(G_{AE},G_{HE}) \\ &+ \lambda_2 \mathcal{L}_{midcyc}(G_{AE})
    \end{split}
    \label{eqn:cyclepath-loss}
\end{equation}

\begin{center}
    \small
    Where $\lambda_1$ and $\lambda_1$ are adjustable coefficients to weight each loss, with $\lambda_1$ set to 10 and $\lambda_2$ set to 50.
\end{center}

The network was trained using the Adam optimiser with a beta 1 value of 0.5 and a cosine-decay learning rate scheduler with an initial value of 1e-3, decreasing to 0 by the end of training.

Our best performing network was trained for 200 epochs, with a batch size of 8. The training dataset consisted of 93,119 pairs of H\&E / AE1/AE3 patches. This was divided into an 80/10/10\% training, validation and test split.
The final network selection was based on a combination of the lowest overall loss as given by equation \ref{eqn:cyclepath-loss}, the highest Staining Dice Coefficient and the lowest Fr\'echet Inception distance.

\newpage

\subsection{Analysis}

To assess the performance of our virtual IHC networks, we had to push beyond a solely visual assessment and adopt methods to generate quantitive values. These would provide concrete evidence of improvement in inference based on alterations to the loss function or network hyperparameters. We developed a method to assess the accuracy of a virtual stain based on stain deconvolution and the Sorensen Dice Coefficient, which we termed the Staining Dice Coefficient. The second method was the Fr\'echet Inception distance, which is widely used in the assessment of generative networks \cite{heusel_gans_2018}. We will describe our implementation of these two methods in the following section.

\subsubsection{The Staining Dice Coefficient (SDC)}

We required a way to quantitatively assess the accuracy of the virtual stains produced by our networks. We speculated that the best method to evaluate the stain reproduction was to separate the stains digitally and then assess how each stain compares to the target image. To separate the stains required colour deconvolution. There were existing methods for stain separation, often used for stain normalisation, such as those discussed in section \ref{sec:introduction}. We settled on the Vahadane structure-preserving colour normalisation method \cite{vahadane_structure-preserving_2016}; this method uses sparse non-negative matrix factorisation to determine a matrix of stain components in optical density space. The inverse of this can be computed and used to calculate staining intensity, converting the RGB image to a stain colourspace. This method was ideal for our needs. It permitted us to separate the RGB images into their stain components both in the real and virtual images and then use this stain colourspace to determine its correlation with the ground truth and, therefore, the accuracy of the virtual stain. 

We made several amendments to the Vahadane method mainly to improve reliability over various slide types and avoid artefacts affecting the resulting matrix; however, one significant difference is that we only used non-negative rather than sparse non-negative factorisation. This was because Vahadane uses sparse factorisation as they conclude that a pixel either has one stain or the other. However, we propose that it more accurately reflects the true biology if the tissue could have components of both stains at any one pixel. We also sampled pixels from across the whole slide to compute our stain matrix rather than at just a few points. We did this by computing a tissue mask using a low-resolution image of the slide in LAB colour space. We could then subsample pixels over all of the tissue, and it became more reliable to compute the stain matrix on slides with unusual artefacts. This method could decompose the RGB whole slide images into a stain colour space, where each channel represents the intensity of staining at each pixel.

Once we had a reliable method to separate the slides into their component stains, we used this to create an evaluation metric for our virtual immunohistochemical images, the Staining Dice Coefficient (SDC). The SDC of a virtual slide was calculated by thresholding the staining intensity in each of the AE1/AE3 and Haematoxylin channels. We repeated this for the target and the virtual IHC images. These binary masks were suitable for calculating a Sorensen-Dice coefficient, often termed the ``Dice Coefficient'' for each stain. This statistic is often used to gauge the correlation between two samples. We intend to represent the accuracy of the overlap of the virtual stain with the ground truth. The staining dice coefficient was computed using the binary masks of staining intensity along with the formula for the Sorensen-Dice Coefficient given in equation \ref{eqn:dice-loss}. X is the binary mask of a stain in the target slide, and Y is the binary mask of the staining intensity in the virtual slide.

\begin{equation}
    \text{Dice Coefficient} = \frac{ 2 | X \cap Y | }{ | X | + | Y | }
    \label{eqn:dice-loss}
\end{equation}

\newpage
\subsubsection{Fr\'echet Inception distance}
\label{sec:frechet}

The Staining Dice Coefficient evaluated the accuracy of overlap of stained areas. Still, we needed a second metric to scrutinise the features and, therefore, the structural detail of the generated images. The Fr\'echet Inception distance (FID) is a widely used metric for deep generative networks that fits this requirement. It was initially proposed by Heusel et al. in 2017\cite{heusel_gans_2018}. 

The FID uses the Inception V3 model during computation, with pre-trained weights from the imagenet dataset. The Inception model is loaded, and the output classification layer is removed, and this leaves the last global pooling layer as an output feature vector. This output is a vector of length 2048, and it is used to capture the features of an input image. 

This configuration can then be employed to evaluate the quality of a generated image. A 2048 long feature vector is predicted for a collection of real images from the target domain to provide a reference for the features of authentic images. Next, a collection of feature vectors is calculated for the generated images, resulting in two groups of feature vectors, one for genuine and one for generated images.

The FID score can then be calculated to measure the effective distance between the distributions of the two collections. The score is calculated using the formula given by Heusel et al. shown in equation \ref{eqn:fid}.

\begin{equation}
    d^{2} = | \mu_1 - \mu_2 |^2 + Tr(\sigma_1 + \sigma_2 - 2*sqrt(\sigma_1*\sigma_2))
    \label{eqn:fid}
\end{equation}

The FID score is referred to as $d^2$ in the paper to show that it is a distance and has squared units\cite{heusel_gans_2018}. $\mu_1$ and $\mu_2$ refer to the feature-wise mean of the real and inferred images. $\sigma_1$ and $\sigma_2$ refer to the covariance matrix for the real and inferred feature vectors. $Tr$ refers to the linear algebra trace operation, which is the sum of elements along the main diagonal of a square matrix.

Now we have a quantitive way to summarise how similar the two collections are in terms of statistics. Where a lower FID score indicates that the two collections of images are more alike or have a more equivalent distribution. A perfect score would be 0.0, indicating that the two collections are identical. Combined with the Staining Dice Coefficient, this provides a method to assess the correctness of stained regions in virtual IHC images and the authenticity of the image features.

\newpage
\section{Results}

\subsection{Dataset Generation}
To assess the usefulness of GANs \cite{goodfellow_generative_2014} in virtual immunohistochemistry of colon tumours, we used a training dataset made up of consecutive pairs of serial sections, with one stained with H\&E and the other with pan-cytokeratin AE1/AE3.  Restained slides, in which the same section is stained first with H\&E then by IHC, were unavailable, so patches were aligned computationally.

We applied a stain normalisation technique over our paired whole slide dataset to mitigate the variations in colour profile and stain intensity present in our training and test datasets. Initial attempts used the Reinhard stain normalisation technique, which moves the mean and standard deviation of channels in the LAB colour space of a source image to match a target image \cite{reinhard_color_2001}. However, when used over entire slide images, this shifts the background values towards the mean, making them discoloured. Therefore, to use the Reinhard method effectively, the tissue must be separated from the background and normalised independently. This can cause image artefacts around the borders of the tissue, and while mainly successful, it is not desirable for deep learning. Therefore we sought a structure-preserving colour normalisation technique.

We found that the Vahadane \cite{vahadane_structure-preserving_2016} technique of structure-preserving colour normalisation was superior to the Reinhard method and incorporated it as our chosen method of colour deconvolution for stain normalisation. However, when used with real-world slides, its described method of patch sampling was susceptible to large artefacts distorting the calculation of the stain matrix and throwing off the colour of the normalised slide. We countered this by calculating the stain matrix based on a subsample of pixels from all tissue in the whole slide image. This allowed for robust automated stain matrix estimation and normalisation across large datasets.

\subsection{GAN Based Virtual IHC}

Figure \ref{fig:gan_example} shows our initial experiments with a generic Generative Adversarial Network (GAN) \cite{goodfellow_generative_2014}. It produced remarkably credible pathology images. The GAN generated images were very similar to the actual tissue and, in isolation, would pass a pathologist's inspection as authentic. However, the inferred slides tended to have areas that were realistic in appearance but that were fictional in terms of staining or cell structure as it would often change the number and shape of cell nuclei between the source and target image. A pathology tool that invents areas of staining or structure is not suitable and indeed dangerous in its realism and the trust that it could therefore instil. For example, small areas of altered tissue could change the resulting diagnosis, especially for processes such as tumour bud scoring, where the presence or absence of a single tumour cell can make a significant difference. Once we identified the inclination of GANs to invent features, it led us to look for alternative network architectures that promoted consistency of tissue in addition to realism.

\subsection{CycleGAN Based Virtual IHC}

\begin{figure}[h]
    \centering
    \includegraphics[width=\textwidth]{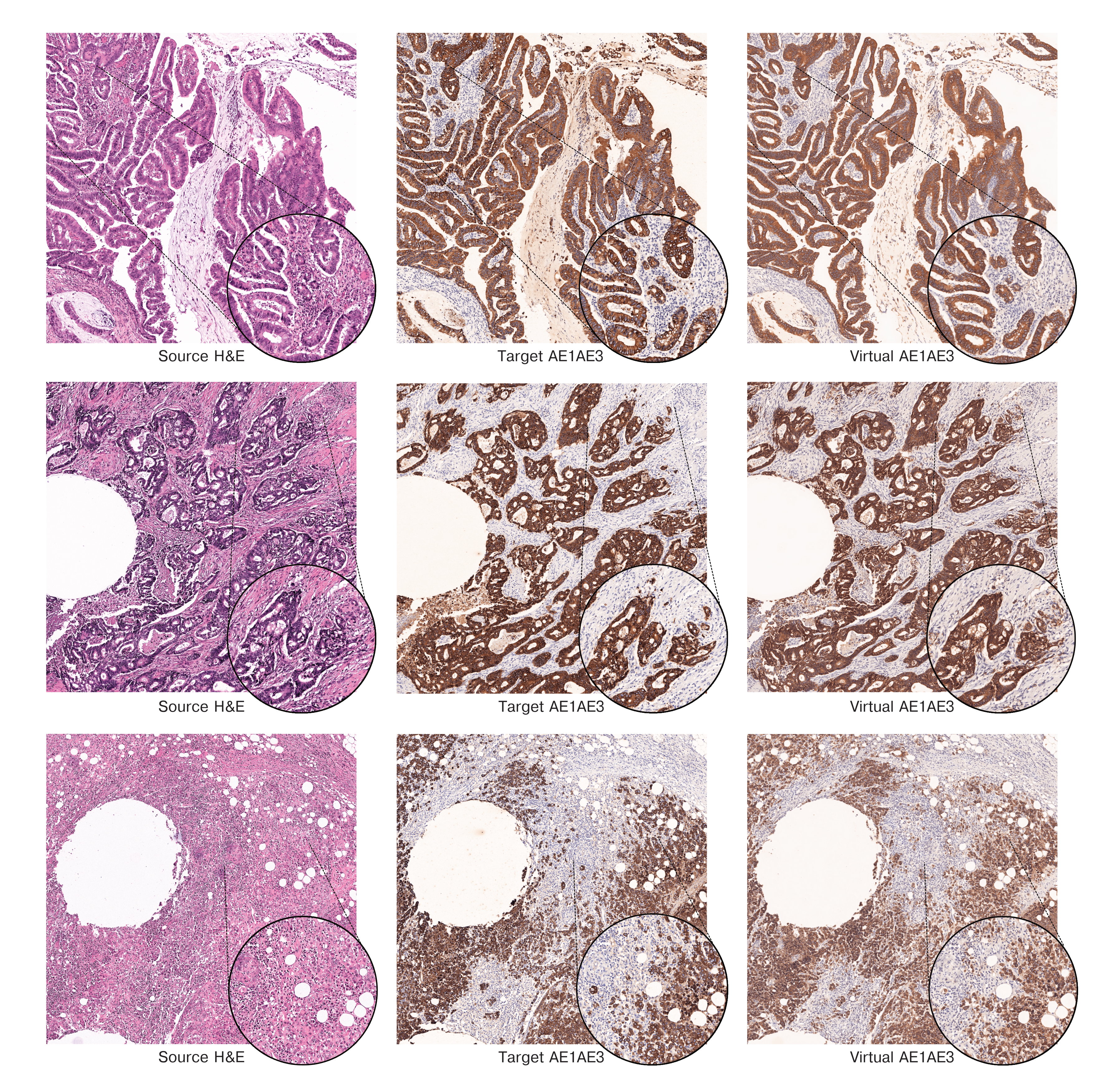}
    \caption{Examples from a CycleGAN with mid-cycle Loss.}
    \label{fig:mid-cycle-examples}
\end{figure}

\begin{figure}[h]
    \centering
    \includegraphics[width=\textwidth]{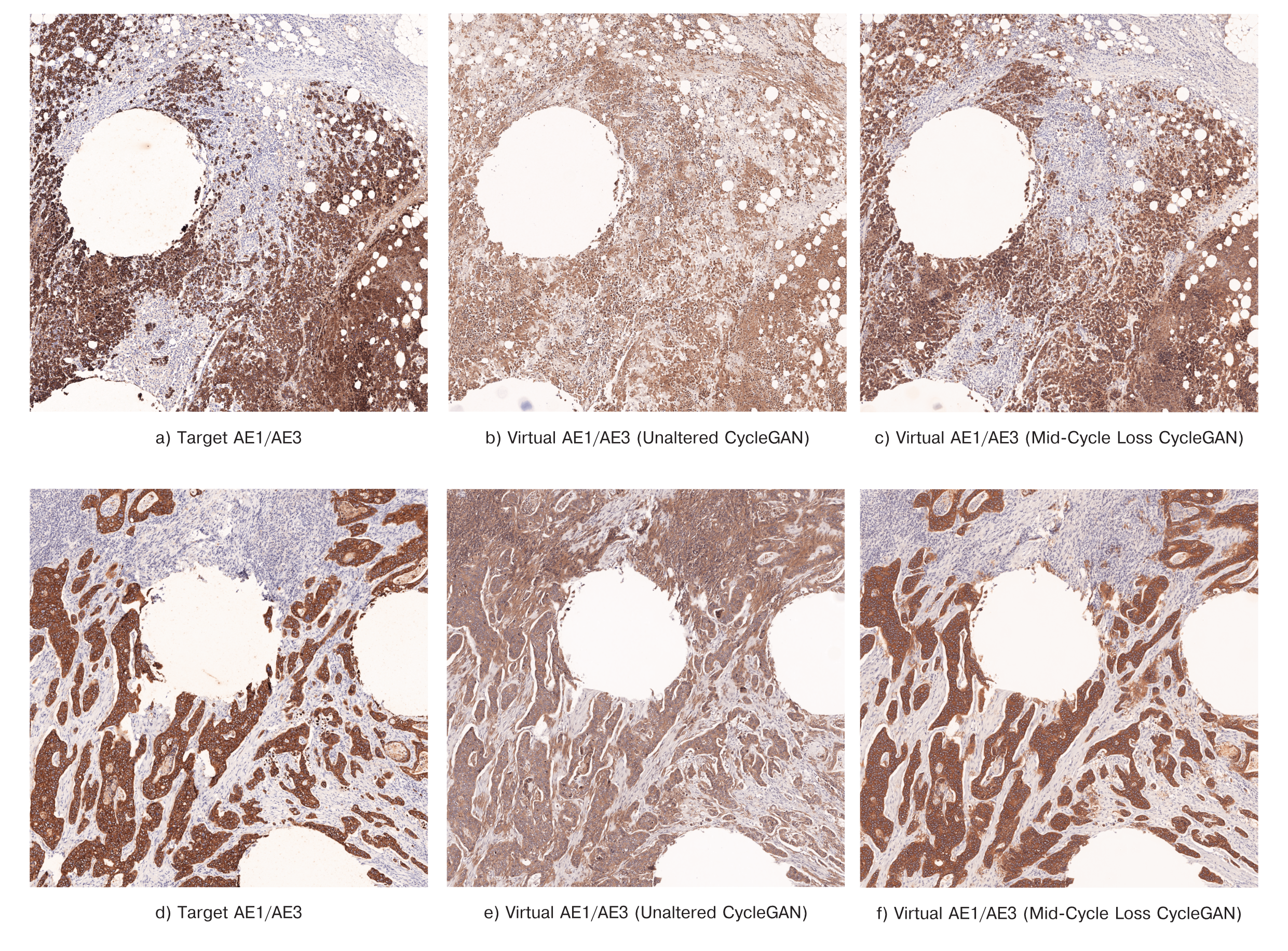}
    \caption{Inference examples from an unaltered CycleGAN and one trained to utilise mid-cycle loss. (Note the 0.6 \textmu m holes resulting from tissue extracted for TMA construction.)}
    \label{fig:cyclegan-comparison}
\end{figure}

The CycleGAN architecture is well suited to meet the requirements for structural consistency. However, cycle loss alone is insufficient to constrain a deep generative network to reproduce the correct staining (Figure \ref{fig:cyclegan-comparison}). A CycleGAN will very accurately recapitulate the cell morphology of input tissue during virtual stain translation. However, they have a similar weakness to GANs in that they can omit or incorrectly fabricate the stain colour. Some examples of this can be seen in figure \ref{fig:cyclegan-comparison}. Here, inaccurate staining produced by an unaltered CycleGAN can be seen in panels (b) and (e) compared to the actual AE1/AE3 stained tissue in panels (a) and (d). Therefore, we identified routes to modify the CycleGAN to reproduce the stain correctly.

\subsection{CycleGAN with Mid-Cycle Loss Based Virtual IHC}

We reasoned that we could constrain the CycleGAN's virtual IHC stain to match the target in a comparable manner to the process for structure preservation. This was achieved by adding a new term to the CycleGAN loss function, designated the "mid-cycle" loss. The mid-cycle loss term modifies the network to compare the generated virtual stain to the ground truth at the mid-point of the training cycle. For it, we chose an error minimisation function that compares overall similarity, mean absolute error. This concentrates on the aggregate of errors across the image, and therefore for this use case, it focuses on the overall stain reproduction. This is successful globally because the cycle loss concentrates on reproducing the structural detail. A weighting term is applied to the structure-preserving cycle loss and stain-preserving mid-cycle loss components. These can be adjusted to fine-tune the network’s prioritisation to become highly effective at H\&E to IHC stain translation while retaining the cellular detail from the source image. This is superior to prior generative virtual stain translation efforts as the blend mitigates the previous limitations of GANs in fabricating cellular structure and the propensity of an unaltered CycleGAN to stain erroneously. Some examples demonstrating the success of virtual IHC patches from a CycleGAN with mid-cycle loss are shown in figure \ref{fig:mid-cycle-examples}.

\begin{figure}[h]
    \centering
    \includegraphics[width=\textwidth]{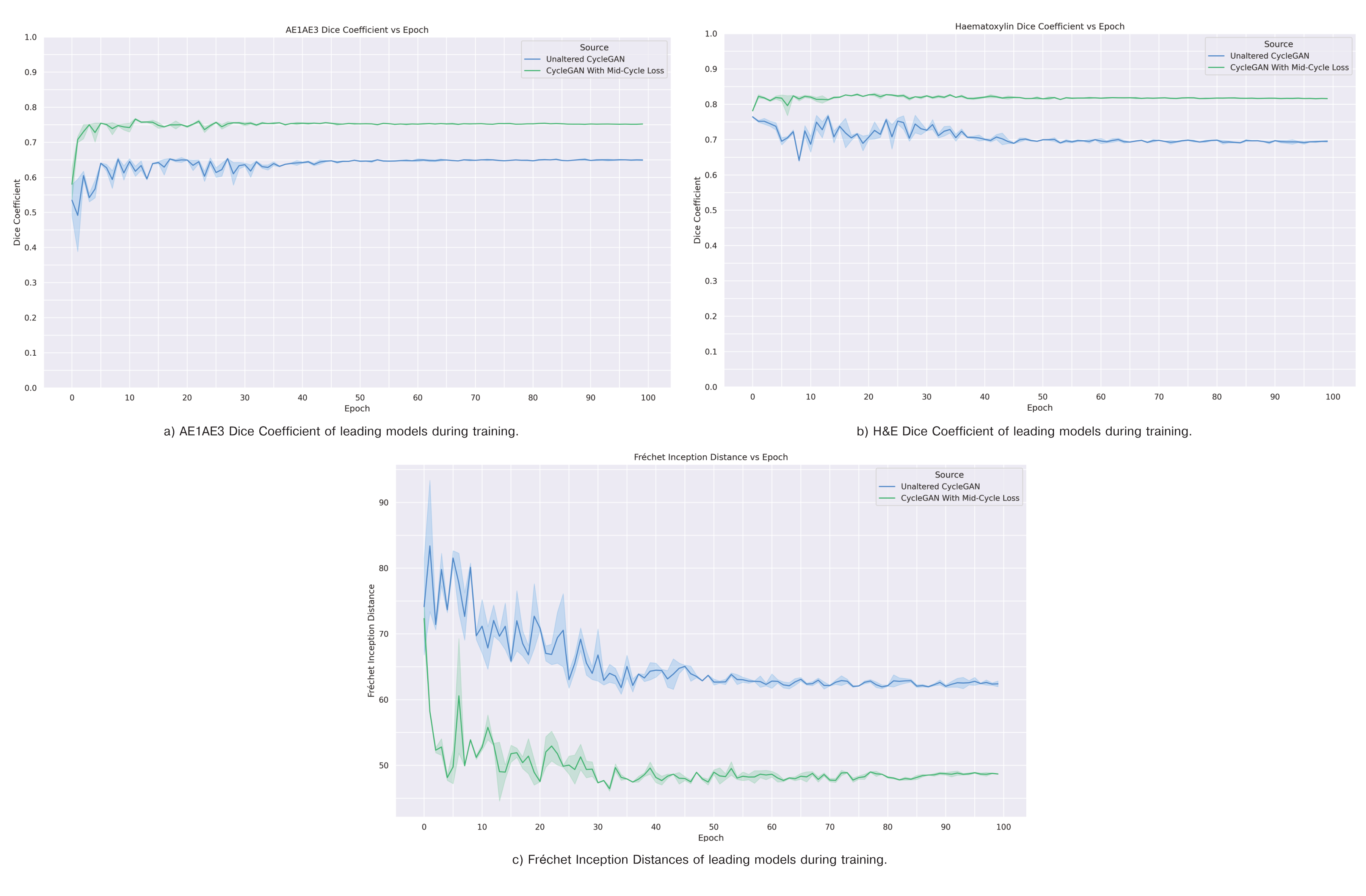}
    \caption{Comparison of training Dice Coefficients and Fr\'echet Inception distances on validation data.}
    \label{fig:training-examples}
\end{figure}

The addition of mid-cycle loss to the CycleGAN improved both the training, validation and test metrics. The CycleGAN with mid-cycle loss produced superior visual results and had lower training and validation losses that were more stable. In addition to these values, we used two other metrics to determine the best model weights for inference.

We held back ten 5000x5000x3 source and target patches for use as validation data. These were already aligned and stain-normalised from our training slides. They were split into smaller patches with a shape relevant to their intended use. For inference in our CycleGAN, they were of shape 512x512x3, and for use with a pre-trained Inception V3 network, they were of shape 299x299x3. These could then be utilised to evaluate the trained models with our chosen validation metrics to select the best epoch for inference.

The first metric of interest was the staining dice coefficient. This was calculated from the binary masks of stained pixels in the inferred and target patches. We generated the masks from a threshold of stain intensity values obtained by deconvolving the RGB images into a stain colour space. This allowed the determination of virtual staining accuracy by comparing the overlap in stained areas in the virtual image and the real target patch.

The second validation metric was the Fr\'echet Inception distance (FID). This was calculated using the following process. First, the source 5000x5000x3 H\&E patches from the validation dataset were input into our CycleGAN, and we generated a virtual AE1/AE3 IHC slide. This was split with the same patch coordinates as our target validation patches. Next, an Inception V3 model was prepared by loading pre-trained weights from the ImageNet dataset. Then the classification head was removed, and the prior 2048 wide pooling layer was used as an output feature vector. This vector can be used to represent the features present in an image. We then used this configuration to compute two collections of feature vectors—one for our real target patches and one for our inferred virtual patches.

The Fr\'echet Inception distance is then calculated from these two collections. The precise method of calculation is given in section \ref{sec:frechet}. However, an important observation is that the closer the two feature vectors are in distribution, the lower the FID and the higher the visual correspondence between the real and virtual patches.

The staining dice coefficient and Fr\'echet Inception distance gave a quantitive way to evaluate the output virtual IHC images and directly compare the performance of an unaltered CycleGAN and one that includes mid-cycle loss during training. In combination, they allow the evaluation of the accuracy of the stained areas and the realism of the output features.

\begin{figure}[h!]
    \centering
    \includegraphics[width=\textwidth]{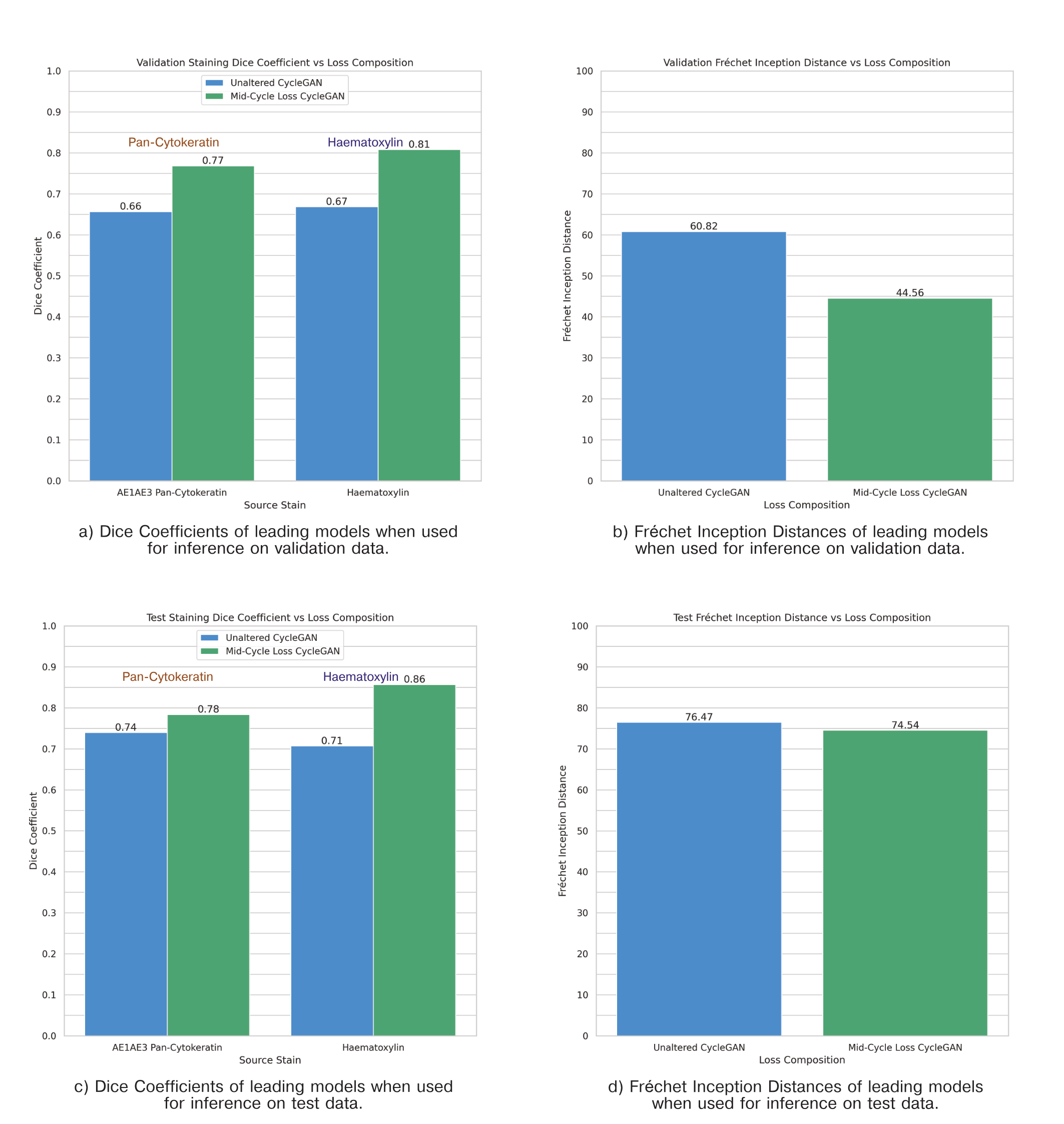}
    \caption{Comparison of stain Dice Coefficients and Fr\'echet Inception distances of leading models on validation and test data.}
    \label{fig:test-comparisons}
\end{figure}

Figure \ref{fig:training-examples} panels (a) and (b) show graphs for the AE1/AE3 and H\&E dice coefficients for the unaltered CycleGAN. The dice coefficients for the CycleGAN with mid-cycle loss are higher for both stains, showing improved accuracy in the overlap of virtual stained areas with the real target tissue. Additionally, the training runs of the CycleGAN with mid-cycle loss were less erratic in improving and reached stability faster than the unaltered CycleGAN, perhaps indicating a more stable and superior training process.

In figure \ref{fig:training-examples} panel (c), a graph of the Fr\'echet Inception distance shows the difference between the virtual and target patches for the unaltered CycleGAN and our CycleGAN with mid-cycle loss. The mid-cycle loss network is again less erratic and quickly reaches a much lower FID. This demonstrates an improved training process and that the resulting features of the mid-cycle loss CycleGAN are closer to the target patches than the unaltered CycleGAN.

Finally, we evaluated our CycleGAN with mid-cycle loss against an unaltered CycleGAN using test data. We created a new dataset from two new pairs of serial sections to ensure that it was entirely unseen data. We assessed the two CycleGANs using the staining dice coefficient and Fr\'echet Inception distance. This allowed us to evaluate the correctness of the stained areas between the virtual and target tissue and the realism of the output features. The resulting test scores on the two leading networks are available in figure \ref{fig:test-comparisons}.

Figure \ref{fig:test-comparisons} panel (c) compares the staining dice coefficients of the virtual slide generated from test data with the corresponding ground truth for each stain component. The staining dice coefficient of the CycleGAN with mid-cycle loss outperforms the unaltered one for both AE1/AE3 and H\&E stains, where the mid-cycle loss CycleGAN has a H\&E dice coefficient of 0.86 vs the unaltered CycleGAN score of 0.71 and an AE1/AE3 dice coefficient of 0.78 vs the unaltered score of 0.74. The Fr\'echet Inception distance between the virtual test slides and ground truth is similarly improved by the addition of mid-cycle loss as can be seen in \ref{fig:test-comparisons} panel (d), where the mid-cycle loss CycleGAN has a distance from the target distribution of 74.54 vs 76.47 for the unaltered CycleGAN.

\section{Discussion}
This work demonstrates that virtual immunohistochemistry is a viable alternative that can successfully recapitulate the characteristics of physical IHC. We have generated large numbers of virtual IHC slides and have confidence in the ability of a deep network to reproduce IHC staining on the tumour core, in small clusters and for well-defined single cells. However, training on consecutive slices might not be the most suitable approach for achieving high accuracy on single tumour cells. Over the 2.5-micron interval of our sections, we observed significant changes in tissue morphology. Nevertheless, the accuracy attained using deformably realigned sections exceeded expectations and is viable when there is no alternative to restaining the same slide.

The implemented modifications to the Vahadane structure-preserving colour normalisation technique performed well and were robust across large datasets; it has advantages in that it is computationally simple enough to run over whole slide images for a large dataset within a reasonable period. Additionally, normalising the input H\&E intensity in the stain colour space ensured that the slides belonged to the same domain without introducing boundary artefacts present in other methods. Normalising the target, AE1/AE3 Pan-Cytokeratin slides produced an idealised output stain that the network could learn to replicate.

Our proposed technique for assessing virtual IHC staining accuracy, the staining dice coefficient, provides access to a quantitive metric to evaluate the effectiveness of deep networks specialised to translate stains for brightfield microscopy. This means that when combined with the Fr\'echet Inception distance, networks can be trained, iterated over, compared and improved with concrete metrics relating to the accuracy of stained areas and the realism of the resulting virtual images. Quantitive metrics for evaluating stain accuracy and structural detail are critical when developing deep networks for pathology. We hope to use these to guide the development of future network design by altering network hyperparameters and observing the effect on these metrics. A tangible value for assessment should help steer hyperparameter selection and improve the accuracy and reliability of deep depth learning applications in virtual immunohistochemistry.

The modifications we have suggested to the CycleGAN loss function exhibit the ability to enforce realistic staining characteristics in a virtual immunohistochemical slide and improve all evaluated metrics compared to an unaltered CycleGAN. Its success lies in calculating the mean absolute error between the generated IHC patch and its corresponding target IHC patch at the mid-point of the cycle. This, in combination with the original CycleGAN cycle consistency loss, can train a network to produce a realistic virtual IHC stain. We hope this will be accurate enough for use in real-world cancer pathology after refinement using improved and expanded datasets. A CycleGAN with our Virtual IHC specific loss function overcomes the tendency of standard GANs to fabricate tissue structures and should allow for near real-time access to immunohistochemistry and the increased contrast that it can provide. Virtual IHC will allow for extensive use, improving the accuracy of diagnosis and unlocking more opportunities for research while reducing the cost and complexity of its adoption.

\section{Data Availability}

All training slides are available at request from the Glasgow Tissue Research Facility and the NHS SafeHaven Biorepository.

\section{Acknowledgements}

We would like to acknowledge and show our appreciation for the work done by Dr A. Ammar, Dr J. Hay, H. Morgan, C. Nixon, Dr J. Quinn, and the staff at the Glasgow Tissue Research Facility for their preparation of the H\&E and IHC slides for the primary and validation datasets, that made this work possible.

We would also like to thank Dr Ke Yuan from the University of Glasgow School of Computing Science for his helpful comments and feedback on the manuscript.

\section{Funding}

This work was funded by CRUK (grant number A24450, to RHI), The Wellcome Trust (grant number 221786/Z/20/Z to RHI), InnovateUK (grant number 42497 to JE), and a studentship to CW from the Glasgow Cancer Centre.

\newpage

\printbibliography

@article{li_automated_2020,
	title = {Automated {Gleason} {Grading} and {Gleason} {Pattern} {Region} {Segmentation} {Based} on {Deep} {Learning} for {Pathological} {Images} of {Prostate} {Cancer}},
	volume = {8},
	issn = {2169-3536},
	doi = {10.1109/ACCESS.2020.3005180},
	journal = {IEEE Access},
	author = {Li, Y. and Huang, M. and Zhang, Y. and Chen, J. and Xu, H. and Wang, G. and Feng, W.},
	year = {2020},
	note = {Conference Name: IEEE Access},
	pages = {117714--117725},
}

@article{langer_computer-aided_2015,
	title = {Computer-aided diagnostics in digital pathology: automated evaluation of early-phase pancreatic cancer in mice},
	volume = {10},
	issn = {1861-6429},
	shorttitle = {Computer-aided diagnostics in digital pathology},
	url = {https://doi.org/10.1007/s11548-014-1122-9},
	doi = {10.1007/s11548-014-1122-9},
	language = {en},
	number = {7},
	urldate = {2021-08-19},
	journal = {International Journal of Computer Assisted Radiology and Surgery},
	author = {Langer, L. and Binenbaum, Y. and Gugel, L. and Amit, M. and Gil, Z. and Dekel, S.},
	month = jul,
	year = {2015},
	pages = {1043--1054},
}

@article{fleming_colorectal_2012,
	title = {Colorectal carcinoma: {Pathologic} aspects},
	volume = {3},
	language = {en},
	number = {3},
	journal = {Journal of Gastrointestinal Oncology},
	author = {Fleming, M. and Ravula, S. and Tatishchev, S.F. and Wang, H.L.},
	year = {2012},
	pages = {21},
}

@article{kim_immunohistochemistry_2016,
	title = {Immunohistochemistry for {Pathologists}: {Protocols}, {Pitfalls}, and {Tips}},
	volume = {50},
	issn = {2383-7837, 2383-7845},
	shorttitle = {Immunohistochemistry for {Pathologists}},
	url = {http://jpatholtm.org/journal/view.php?doi=10.4132/jptm.2016.08.08},
	doi = {10.4132/jptm.2016.08.08},
	language = {en},
	number = {6},
	urldate = {2022-04-06},
	journal = {Journal of Pathology and Translational Medicine},
	author = {Kim, S. and Roh, J. and Park, C.},
	month = nov,
	year = {2016},
	pages = {411--418},
}

@article{duraiyan_applications_2012,
	title = {Applications of immunohistochemistry},
	volume = {4},
	issn = {0976-4879},
	url = {https://www.ncbi.nlm.nih.gov/pmc/articles/PMC3467869/},
	doi = {10.4103/0975-7406.100281},
	number = {Suppl 2},
	urldate = {2022-04-06},
	journal = {Journal of Pharmacy \& Bioallied Sciences},
	author = {Duraiyan, J. and Govindarajan, R. and Kaliyappan, K. and Palanisamy, M.},
	month = aug,
	year = {2012},
	pmid = {23066277},
	pmcid = {PMC3467869},
	pages = {S307--S309},
}

@article{jackson_machine_2020,
	title = {A machine learning algorithm for simulating immunohistochemistry: development of {SOX10} virtual {IHC} and evaluation on primarily melanocytic neoplasms},
	volume = {33},
	issn = {0893-3952, 1530-0285},
	shorttitle = {A machine learning algorithm for simulating immunohistochemistry},
	url = {http://www.nature.com/articles/s41379-020-0526-z},
	doi = {10.1038/s41379-020-0526-z},
	language = {en},
	number = {9},
	urldate = {2021-08-03},
	journal = {Modern Pathology},
	author = {Jackson, C.R. and Sriharan, A. and Vaickus, L.J.},
	month = sep,
	year = {2020},
	pages = {1638--1648},
}

@article{lo_cycle-consistent_2021,
	title = {Cycle-consistent {GAN}-based stain translation of renal pathology images with glomerulus detection application},
	volume = {98},
	issn = {1568-4946},
	url = {https://www.sciencedirect.com/science/article/pii/S1568494620307602},
	doi = {10.1016/j.asoc.2020.106822},
	language = {en},
	urldate = {2021-08-17},
	journal = {Applied Soft Computing},
	author = {Lo, Y. and Chung, I. and Guo, S. and Wen, M. and Juang, C.},
	month = jan,
	year = {2021},
	pages = {106822},
}

@article{burlingame_shift_2020,
	title = {{SHIFT}: speedy histological-to-immunofluorescent translation of a tumor signature enabled by deep learning},
	volume = {10},
	issn = {2045-2322},
	shorttitle = {{SHIFT}},
	url = {http://www.nature.com/articles/s41598-020-74500-3},
	doi = {10.1038/s41598-020-74500-3},
	language = {en},
	number = {1},
	urldate = {2021-08-03},
	journal = {Scientific Reports},
	author = {Burlingame, E.A. and McDonnell, M. and Schau, G.F. and Thibault, G. and Lanciault, C. and Morgan, T. and Johnson, B.E. and Corless, C. and Gray, J.W. and Chang, Y.H.},
	month = dec,
	year = {2020},
	pages = {17507},
}

@article{ronneberger_u-net_2015,
	title = {U-{Net}: {Convolutional} {Networks} for {Biomedical} {Image} {Segmentation}},
	shorttitle = {U-{Net}},
	url = {http://arxiv.org/abs/1505.04597},
	language = {en},
	urldate = {2021-08-03},
	journal = {arXiv:1505.04597 [cs]},
	author = {Ronneberger, O. and Fischer, P. and Brox, T.},
	month = may,
	year = {2015},
	note = {arXiv: 1505.04597},
}

@article{fragemann_review_2022,
	title = {Review of {Disentanglement} {Approaches} for {Medical} {Applications} -- {Towards} {Solving} the {Gordian} {Knot} of {Generative} {Models} in {Healthcare}},
	url = {http://arxiv.org/abs/2203.11132},
	urldate = {2022-03-23},
	journal = {arXiv:2203.11132 [cs]},
	author = {Fragemann, J. and Ardizzone, L. and Egger, J. and Kleesiek, J.},
	month = mar,
	year = {2022},
	note = {arXiv: 2203.11132},
}

@article{gui_review_2020,
	title = {A {Review} on {Generative} {Adversarial} {Networks}: {Algorithms}, {Theory}, and {Applications}},
	shorttitle = {A {Review} on {Generative} {Adversarial} {Networks}},
	url = {https://arxiv.org/abs/2001.06937v1},
	doi = {10.48550/arXiv.2001.06937},
	language = {en},
	urldate = {2022-03-23},
	author = {Gui, J. and Sun, Z. and Wen, Y. and Tao, D. and Ye, J.},
	month = jan,
	year = {2020},
}

@article{goodfellow_generative_2014,
	title = {Generative {Adversarial} {Networks}},
	url = {http://arxiv.org/abs/1406.2661},
	language = {en},
	urldate = {2021-08-03},
	journal = {arXiv:1406.2661 [cs, stat]},
	author = {Goodfellow, I.J. and Pouget-Abadie, J. and Mirza, M. and Xu, B. and Warde-Farley, D. and Ozair, S. and Courville, A. and Bengio, Y.},
	month = jun,
	year = {2014},
	note = {arXiv: 1406.2661},
}

@article{saxena_comparison_2021,
	title = {Comparison and {Analysis} of {Image}-to-{Image} {Generative} {Adversarial} {Networks}: {A} {Survey}},
	shorttitle = {Comparison and {Analysis} of {Image}-to-{Image} {Generative} {Adversarial} {Networks}},
	url = {http://arxiv.org/abs/2112.12625},
	urldate = {2022-03-23},
	journal = {arXiv:2112.12625 [cs]},
	author = {Saxena, S. and Teli, M.N.},
	month = dec,
	year = {2021},
	note = {arXiv: 2112.12625},
}

@article{zhu_unpaired_2020,
	title = {Unpaired {Image}-to-{Image} {Translation} using {Cycle}-{Consistent} {Adversarial} {Networks}},
	url = {http://arxiv.org/abs/1703.10593},
	language = {en},
	urldate = {2021-08-03},
	journal = {arXiv:1703.10593 [cs]},
	author = {Zhu, J.Y. and Park, T. and Isola, P. and Efros, A.A.},
	month = aug,
	year = {2020},
	note = {arXiv: 1703.10593},
}

@inproceedings{macenko_method_2009,
	address = {Boston, MA, USA},
	title = {A method for normalizing histology slides for quantitative analysis},
	isbn = {978-1-4244-3931-7},
	url = {http://ieeexplore.ieee.org/document/5193250/},
	doi = {10.1109/ISBI.2009.5193250},
	language = {en},
	urldate = {2022-02-09},
	booktitle = {2009 {IEEE} {International} {Symposium} on {Biomedical} {Imaging}: {From} {Nano} to {Macro}},
	publisher = {IEEE},
	author = {Macenko, M. and Niethammer, M. and Marron, J. S. and Borland, D.and Woosley, J.T. and Xiaojun, G. and Schmitt, C. and Thomas, N.E.},
	month = jun,
	year = {2009},
	pages = {1107--1110},
}

@article{reinhard_color_2001,
	title = {Color {Transfer} between {Images}},
	language = {en},
	journal = {IEEE Computer Graphics and Applications},
	author = {Reinhard, E. and Ashikhmin, M. and Gooch, B. and Shirley, P.},
	year = {2001},
	pages = {8},
}

@article{vahadane_structure-preserving_2016,
	title = {Structure-{Preserving} {Color} {Normalization} and {Sparse} {Stain} {Separation} for {Histological} {Images}},
	volume = {35},
	issn = {1558-254X},
	doi = {10.1109/TMI.2016.2529665},
	number = {8},
	journal = {IEEE Transactions on Medical Imaging},
	author = {Vahadane, A. and Peng, T. and Sethi, A. and Albarqouni, S. and Wang, L. and Baust, M. and Steiger, K. and Schlitter, A.M. and Esposito, I. and Navab, N.},
	month = aug,
	year = {2016},
	note = {Conference Name: IEEE Transactions on Medical Imaging},
	pages = {1962--1971},
}

@article{heusel_gans_2018,
	title = {{GANs} {Trained} by a {Two} {Time}-{Scale} {Update} {Rule} {Converge} to a {Local} {Nash} {Equilibrium}},
	url = {http://arxiv.org/abs/1706.08500},
	language = {en},
	urldate = {2022-02-09},
	journal = {arXiv:1706.08500 [cs, stat]},
	author = {Heusel, M. and Ramsauer, H. and Unterthiner, T. and Nessler, B. and Hochreiter, S.},
	month = jan,
	year = {2018},
	note = {arXiv: 1706.08500},
}






\end{document}